# An ISAC-ready Full-Duplex Backscatter Architecture for the mmWave IoT

Skanda Harisha, Jimmy G. D. Hester, Aline Eid
{skandah,jghester,alineeid}@umich.edu
University of Michigan, Ann Arbor, USA

## Abstract

Achieving long-range, high data rate, concurrent two-way mmWave communication with power-constrained IoT devices is fundamental to scaling future ubiquitous sensing systems, yet the substantial power demands and high cost of mmWave hardware have long stood in the way of practical deployment. This paper presents Armstrong, the first mmWave full-duplex backscatter tag architecture, charting a genuinely low-cost path toward high-performance mmWave connectivity for ISAC systems. Armstrong operates in full duplex at ranges beyond 88$m$ and beyond 200$m$ in downlink alone, delivering 20× the reach of state-of-the-art systems while being over 100× cheaper than existing mmWave backscatter platforms. Enabling this leap is a novel low-power regenerative amplifier that provides 30 dB of gain while consuming only 7.7 mW during active transmission, paired with a regenerative rectifier that achieves state-of-the-art sensitivity down to −60 dBm. We integrate our circuits on a compact PCB and evaluate it across diverse downlink and uplink scenarios, where it achieves 1 Kbps BERs of less than $10^{-1}$ at 200$m$ and 88$m$, respectively, demonstrating resilient, high-quality communication even at extended ranges.

## Keywords

Full Duplex, Backscatter communication, 5G/6G, millimeter wave, mmWave Tags

## 1 Introduction

The modern zeitgeist around next-generation (6G) cellular networks is characterized by a shift away from a sole focus on communications between base stations and handheld devices, and towards a more holistic and multifaceted concern for communications between and sensing by motley systems ranging from base stations to vehicles and low-cost Internet of Things (IoT) type devices [1]. In the midst of this trend has emerged a keen interest in the concept of Integrated Sensing and Communications (ISAC), where (mostly) base stations are enabled with traditional communication capabilities, on top of which is overlain—with, ideally, little capacity overhead—the ability to spatially image and understand their environments [2–4]. mmWave and terahertz frequency bands have predominantly been identified as the most promising media for ISAC, due to their ability to host electrically large antenna systems and large absolute operating bandwidths—yielding high angular and radial imaging resolutions, respectively. Unfortunately, active base stations have been shown unable to provide sub-decameter User Equipment (UE) localization accuracy [5] in the absence of a high overhead synchronization. Even in ideal conditions, this fragile synchronization is still too tenuous to enable better than 1m of ranging accuracy [6]. These problems are inherent to active communication systems, where the transmitter and receiver are not coherent. Recent efforts have demonstrated the ability of relatively simple and low-cost mmWave systems to provide the tracking of backscatter tags at long ranges with cm-level accuracy [7–10]. This ability is uniquely enabled by the coherence between the signals transmitted by the BS and the location-information-carrying signal backscattered by the UE. However, these systems fail at the other intent of ISAC: communications. While uplink communications—from the UE to the BS—have been the bread and butter of backscatter systems since their inception, backscatter tags are not as well suited to receive downlink streams, due to their lack of the local oscillators required by down-converting mixers. Like commercial Off-The-Shelf (OTS) EPC Gen2 RFIDs, recent efforts have proposed the use of rectifier-based receivers to enable this downlink capability in mmWave backscatter systems [11, 12]—however, using costly dedicated mmWave components, requiring a time division of transmissions and receptions, and with ranges far inferior to that required by 6G ISAC systems. For widespread adoption and seamless integration into existing wireless infrastructures, future mmWave backscatter systems must be compatible with conventional mmWave base stations and support simultaneous bidirectional (full-duplex) communication without the need for specialized hardware.

Full-duplex backscatter tags refers to devices that can receive incoming signals and transmit information concurrently, enabling true bidirectional communication without the need to generate an active radio carrier wave. This capability becomes especially important in many modern day applications such as smart factory asset monitoring, wearable and implantable health sensors, and large-scale Internet of Things (IoT) deployments. In such scenarios, devices must continuously exchange information with minimal delay while operating under stringent energy constraints and at extended ranges. Leveraging millimeter-wave (mmWave) frequencies for backscatter communication further enhances these systems by enabling high data rates, compact antenna designs, and focused beams, which are essential for dense and dynamic environments. However, traditional half-duplex backscatter systems, which separate reception and transmission in time, often fail to meet these requirements due to increased latency. By enabling simultaneous bidirectional communication in the mmWave band, full-duplex backscatter systems can significantly improve throughput, responsiveness, and energy efficiency making them a key enabler for next-generation low-power, high-capacity wireless communication networks.

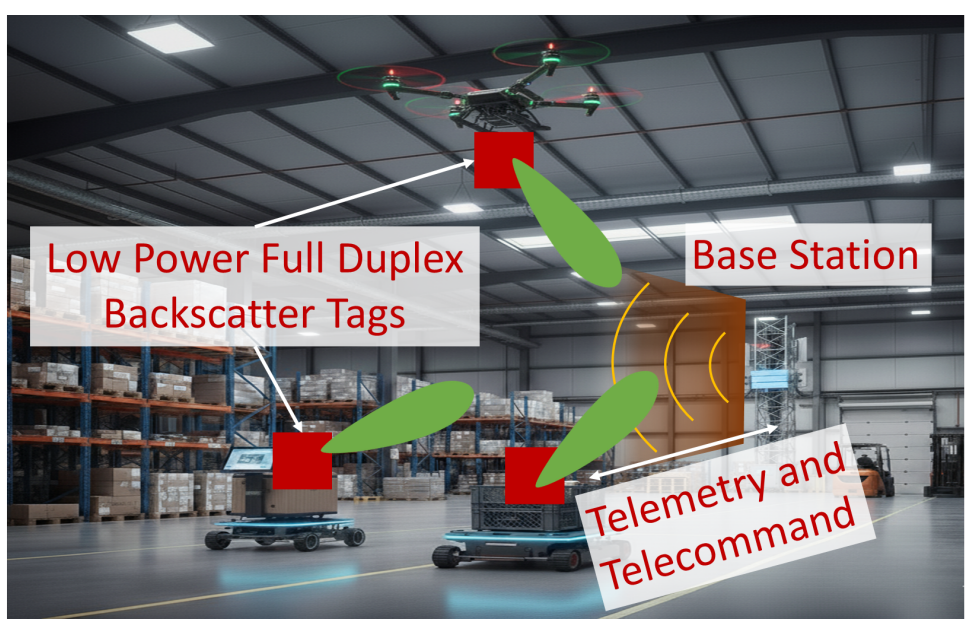

**Figure 1:** Armstrong is a mmWave backscatter tag that is capable of operating in full duplex mode at extended ranges while being interrogated by future ISAC base-stations (Image generated with the assistance of Gemini AI).

In this work, we resurrect and modernize a defining wireless technology of the 1910s to introduce Armstrong, a low-power, fully integrated full-duplex mmWave backscatter tag designed to interface with next generation wireless networks. The core idea behind Armstrong is a dual regenerative architecture that supports concurrent transmission and reception using two complementary components: a regenerative amplifier for uplink communication and a regenerative rectifier for downlink reception. This design enables both communication paths to coexist efficiently within the same hardware while maintaining high signal integrity and low interference. The high-Q nature of both uplink and downlink channels allows multiple tags to share the spectrum within a narrow bandwidth, greatly improving scalability and spectral efficiency. To keep the system simple and energy-efficient, Armstrong uses amplitude shift keying (ASK) for reader-to-tag communication and frequency shift keying (FSK) for tag-to-reader communication. Together, these innovations allow Armstrong to achieve true simultaneous bidirectional communication at mmWave frequencies—bridging the gap between existing half-duplex designs and future large-scale, infrastructure-compatible backscatter networks.

Before outlining our novel system architecture, it is helpful to first highlight the key challenges that must be addressed to realize long-range, full-duplex mmWave backscatter tags:

**Long-range communications:** One of the major advantages of mmWave systems is the ability to use very small front-end antennas, making it easy to integrate them into compact sensors and microcontrollers. However, to compensate for severe path loss at mmWave frequencies, the antenna aperture is typically increased. While a larger aperture provides higher gain, it also narrows the tag's angular coverage. To break this gain-coverage tradeoff, prior works incorporate additional planar structures such as Van Atta arrays [11, 13] or passive true-time-delay beamforming networks such as Rotman lenses [9, 14, 15], and even non-planar structures such as dielectric lenses used in [10]. Although these techniques improve link budget, they make the tag bulky and undermine the original promise of small, unobtrusive IoT devices. This raises an important question: how can we offset mmWave path loss without physically enlarging the antenna aperture? A straightforward idea is to add amplifiers into the RF chain. However, conventional mmWave amplifiers are extremely power-hungry and expensive, making them unsuitable for low-power low-cost IoT systems. This motivates a deeper exploration: can we redesign a mmWave amplifier from the ground up-one optimized specifically for the power and cost constraints of IoT backscatter tags?

**Two-way communication:** State of the art mmWave backscatter tags typically communicate using a high frequency switch, either off-the-shelf [8] or custom built [10, 16] - to alternate between two impedance states with different reflection coefficients. This architecture works well when the primary goal is uplink-only communication, such as transmitting the tag ID. However, in works such as [11, 12], the authors attempt to extend this hardware architecture to support two-way communication with the reader. Their design connects one port of the switch to the downlink decoder and the other port to a standard 50Ω load. During downlink reception, the switch must remain connected to the decoder to avoid losing data, whereas during uplink transmission, it must toggle between the two impedance states, as in conventional backscatter. This configuration inherently restricts the tag to half-duplex operation, since the switch cannot simultaneously provide reliable downlink decoding and controlled uplink modulation. This raises a natural question: how can we design a new tag architecture that overcomes this fundamental limitation and enables true full-duplex backscatter communication?

Armstrong directly addresses these questions by providing a unified solution to both full duplex communication and long range operation in mmWave backscatter systems. In the following, we summarize the key contributions of this work:

• **A full-duplex mmWave backscatter system:** We present the first fully integrated full-duplex mmWave backscatter

system capable of simultaneous high-rate uplink and downlink communication between a tag and a reader. Unlike prior half-duplex designs that require time division for transmission and reception, our architecture enables continuous bidirectional communications. This breakthrough opens up new opportunities for real-time sensing, control, and interactive applications such as wireless AR/VR peripherals, low-latency robotic control, multi-tag collaborative localization, and closed-loop industrial monitoring, all the while maintaining the low-power and low-cost advantages of backscatter communications.

• **A regenerative uplink amplifier for extended range:** We design and integrate a high-Q, low-power regenerative amplifier within the tag to boost uplink signal strength without the need for costly high-power active RF chains. The amplifier employs positive feedback to achieve high gain with minimal power—consuming only $7mW$, which is more than 10× lower than commercial amplifiers of similar gains while costing over 300× less. This enables robust uplink transmission over longer distances, improving SNR and communication reliability.

• **A regenerative rectifier front-end for sensitive downlink reception:** We introduce a novel regenerative rectifier front-end that significantly enhances downlink reception sensitivity. By integrating rectification and regenerative feedback, our design can decode weak incident signals at power levels up to -60$dBm$, offering an improvement of more than 30$dB$ over state-of-the-art commercial active mmWave rectifiers at over 100x lower cost. This improvement enables reliable operation under low illumination, going hand in hand with the uplink channel.

• **System-level integration and experimental validation:** We implement and experimentally evaluate the complete tag–reader prototype, demonstrating true simultaneous full-duplex operation at mmWave frequencies. Our system achieves high Kbps bidirectional data rates, can support multiple concurrent tags within a narrow bandwidth due to its high-Q architecture, and operates reliably at distances exceeding tens of meters under realistic environmental conditions, including NLoS and dynamic settings.

## 2 Background on Regenerative Amplifiers

The year is 1912. A couple thousand radio systems are operating on US soil—more than 80% of which are operated by radio amateurs (a.k.a. Hams) [17]. Most of their receivers utilize a then mysterious device called the crystal (or cat whisker) detector, which would later be identified as none other than a Schottky diode rectifier. Six years earlier, Lee de Forest invented the Audion—the first (poor) vacuum tube and the first electronic amplifying component. However, it was costly and displayed poor gain, leading to the need to cascade several of them to achieve acceptable performance, albeit at high cost. It is then that a Columbia University undergraduate student, Edwin Armstrong, realized that by carefully feeding back the output of an Audeon-based amplifier into its input, much greater gains could be achieved at little additional cost. This architecture—known as the regenerative amplifier—would go on to enhance radio receivers for the next decade, until it was supplanted by superheterodyne architectures. For the mmWave IoT devices considered in this paper, the modern context is similar. MmWave active radios or high-gain amplifiers—requiring complex signal chains built on large areas of expensive GaAs wafers—do not provide a cost-effective solution for such receivers. Enter the regenerative amplifier.

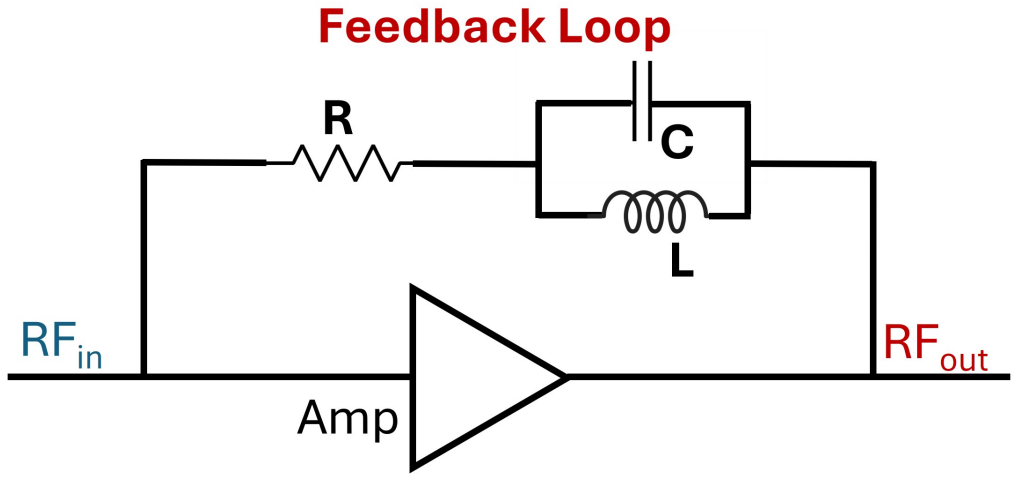

**Figure 2:** The basic schematic of a regenerative amplifier consisting of an amplifier with a positive feedback loop.

Regenerative amplifiers are a fascinating class of circuits that exploit positive feedback to achieve very high gains. They have historically played an important role in radio-frequency (RF) receivers, where they were used to dramatically enhance sensitivity while maintaining relatively low power consumption. At their core, regenerative amplifiers consist of an amplifier combined with a positive feedback network, as shown in Fig. 2. From a control-systems perspective, the closed-loop transfer function of such a system can be written as

$$G'(s) = \frac{G(s)}{1 - f * G(s)} \tag{1}$$

where $G(s)$ is the open-loop gain, $f$ is the feedback factor, and their product is the loop gain. This highlights the key role of the loop gain in determining the system's behavior. A natural question arises: what happens when the loop gain reaches unity?

This condition leads directly to the Barkhausen criterion for oscillation: if the magnitude of the loop gain is unity and the feedback introduces a net phase shift of 360°, the circuit transitions from amplification to sustained oscillation. However, in regenerative amplifiers, we deliberately avoid this condition by carefully tuning the feedback factor. Instead, the system is designed such that the loop gain approaches, but does not reach, unity. In this regime, the amplifier achieves very large effective gain without self-oscillating, enabling significant signal amplification at modest cost and power levels.

Regenerative amplification can also be understood through the *negative-resistance perspective*. Positive feedback effectively introduces a negative resistance that counteracts the inherent ohmic losses of the circuit. As the magnitude of this negative resistance approaches the physical resistance, the overall losses decrease, leading to stronger resonance and higher amplification. This perspective naturally explains why regenerative amplifiers can achieve both high gain and sharp frequency selectivity.

Operating near the oscillation threshold also enhances the circuit's quality factor, given by

$$Q = \frac{X}{R - |R_r|}, \tag{2}$$

where $X$ is the total reactance, $R$ is the ohmic resistance, and $R_r$ represents the effective negative resistance introduced by the feedback. As $|R_r|$ approaches $R$, the denominator becomes small, increasing both the quality factor and the overall gain. This shows that the loop gain, negative resistance, and quality factor are intimately connected in achieving high-performance amplification.

Another key aspect is that the loop gain is inherently frequency dependent, which makes the amplifier naturally selective and can allow multiple amplifiers to operate side-by-side at even slightly different resonant frequencies. In practice, the feedback network is typically implemented using an $LC$ resonant circuit arranged in series or parallel with some resistance. The resonator oscillates at its natural frequency $\omega_0 = 1/\sqrt{LC}$, providing strong frequency discrimination. At resonance, the feedback signal is reinforced most effectively, pushing the loop gain closer to unity, whereas away from resonance the feedback diminishes, reducing the effective gain. The resistance in the $RLC$ network also plays a crucial role: it sets the baseline losses that the negative resistance introduced by feedback must overcome for regeneration to occur. At high frequencies, such $RLC$ networks can be realized using the distributed properties of transmission lines, which naturally exhibit inductive and capacitive behavior. By carefully tuning these transmission-line sections, the circuit can be made frequency selective precisely at the desired frequency - an essential requirement in RF applications to amplify weak signals while rejecting out-of-band noise and interference.

## 3 System Overview

The proposed system consists of full-duplex tags that are interrogated by a custom mmWave reader. Downlink communications use simple Amplitude Shift Keying (ASK), which can be readily supported by most mmWave reader hardware. In parallel, the reader also transmits a continuous single-tone sine wave. The tag receives this tone and embeds its data bits onto it, enabling uplink transmission without needing to generate its own carrier. This uplink employs standard FSK modulation, and the resulting modulated backscatter signal is decoded at the reader using an energy-efficient Goertzel algorithm. A high-level overview of the system is shown in Fig. 3. Each component of the system is described in detail in the following subsections.

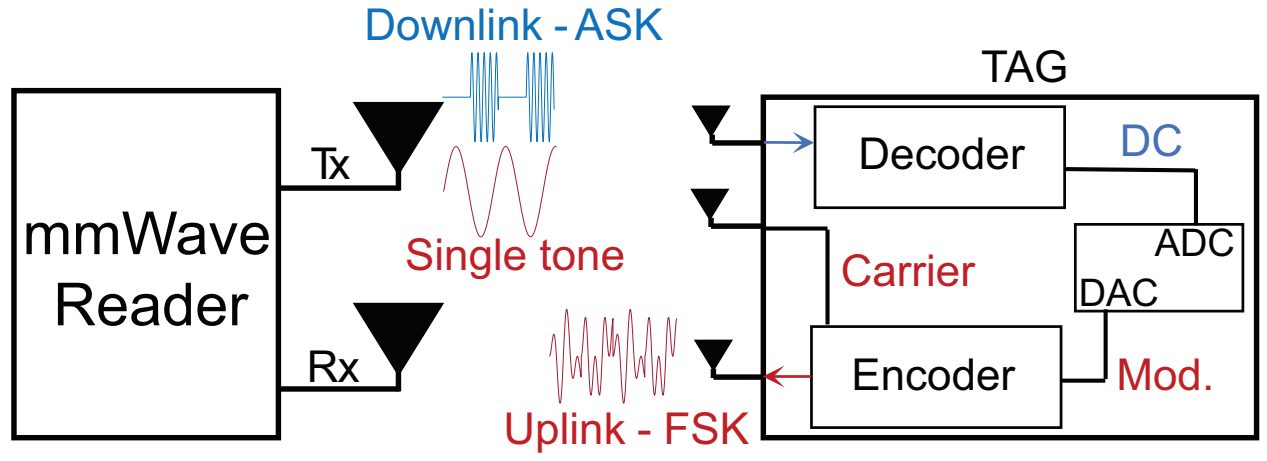

**Figure 3:** The reader transmits an ASK-modulated signal for downlink communication along with a continuous single-tone carrier for backscatter uplink, while the tag operates in full duplex.

### 3.1 mmWave Reader:

For the reader, we use the Sivers EVK02001 Evaluation Kit [18], which provides a wide operating range from 24-29.5$GHz$. The platform supports MIMO operation and shares a common reference clock across its transmit and receive chains, ensuring stable frequency alignment—an essential requirement for coherent mmWave backscatter. Importantly, the EVK02001 supports full-duplex operation, enabling simultaneous transmission and reception on the same hardware. This capability allows a single reader to both interrogate the tag and capture the backscattered uplink signal, significantly simplifying the overall system architecture. At best, the EVK02001 is capable of transmitting 35$dBm$ effective isotropic radiated power (EIRP). The mmWave front end is driven by a Xilinx RFSoC 4×2 board [19], which provides the baseband signals for the reader. The RFSoC includes high-performance DACs that can run up to 9.85 GSPS and ADCs up to 5 GSPS. These data converters support correspondingly wide analog bandwidth: the Gen 3 RFSoC devices (such as on the 4×2 board) have RF-ADC input bandwidths up to 6 GHz. This high sampling rate and broad analog bandwidth allow us to generate and digitize baseband waveforms that cover the full spectrum required for our downlink ASK modulation as well as the single-tone uplink backscatter signal, without the need for external up/down conversion stages.

### 3.2 Armstrong Tag Design

Here we describe the design principles of the various components in the Armstrong tag, as illustrated in the schematic in Fig. 4. The tag uses a broadband antenna to accommodate small shifts in circuit resonance—an important requirement

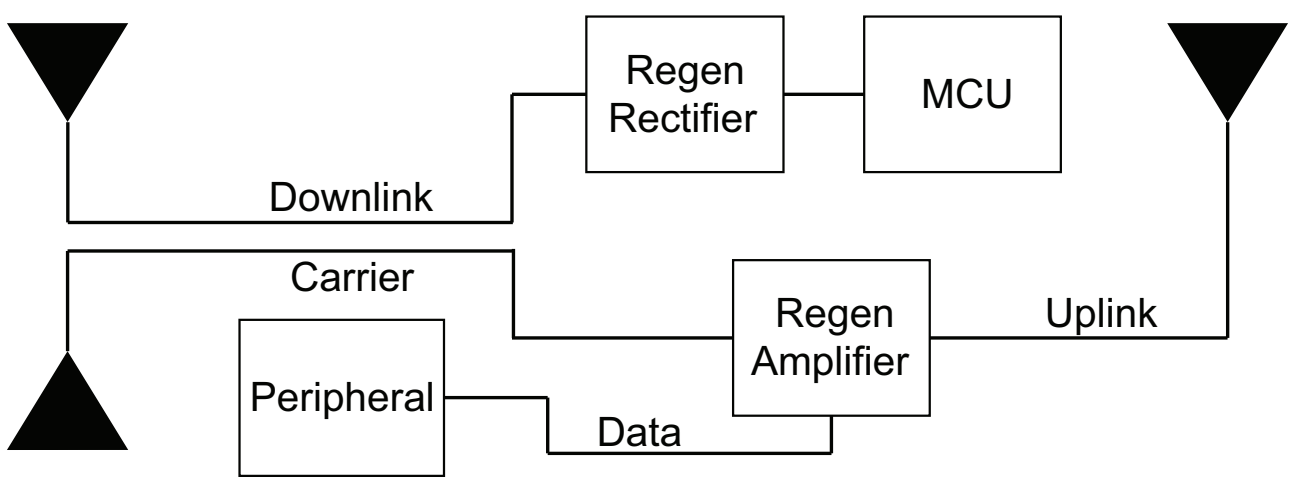

**Figure 4:** Armstrong full-duplex tag architecture.

whose significance will become clear in later sections.

One receiving antenna is connected to the decoder chain, which includes our novel regenerative receiver: a regenerative amplifier precisely tuned to the impedance of a rectifier. This combined architecture enables very high downlink sensitivity while consuming low power. The rectifier output is then delivered to an ultra-low-power microcontroller (MCU) for decoding the received signal.

The second receiving antenna is routed to a standalone regenerative amplifier responsible for uplink communication. This amplifier boosts the incident signal, applies the tag's data modulation, and drives a separate antenna that transmits the enhanced backscatter signal back to the reader.

*3.2.1* ***Wideband Antenna Design:*** Having outlined the overall tag architecture, we now focus on the antenna design, which must operate reliably despite small frequency drifts introduced by the regenerative circuits. This requirement motivates our use of a wideband U-slot microstrip patch antenna with significantly enhanced bandwidth compared to standard mmWave patches. It is advantageous for our system to operate over a wide frequency range so that we can support a larger number of tags without modifying the front-end antennas. Moreover, the resonant frequencies of the regenerative amplifiers and rectifiers can drift slightly due to temperature variations or environmental changes. To ensure robust operation under these conditions, we leveraged the rich literature on wideband patch antenna design [20–22] to build a wideband U-slot microstrip patch with approximately 12% fractional impedance bandwidth—substantially higher than the 2–5% bandwidth typical of conventional patch antennas. This enhanced bandwidth arises from the additional resonant mode introduced by the U-slot. However, at mmWave frequencies, antenna dimensions become extremely sensitive to even small geometric variations, requiring careful optimization. So after the initial preliminary design, we employed particle swarm optimization for broad parameter exploration, followed by the Nelder–Mead algorithm for local refinement to further enhance the gain and impedance bandwidths of the antenna. The final antenna element design seen in Fig.5a, was fully modeled and validated using ANSYS HFSS. To further increase the passive gain of the tag, we constructed a four-element antenna array, with the individual elements combined through Wilkinson power dividers. Each divider provides approximately −3 dB insertion loss from the input to each output branch while maintaining about 20 dB isolation between the two output ports. This high isolation is achieved using a high-frequency 100Ω isolation resistor, ensuring stable performance and minimal coupling between array elements.

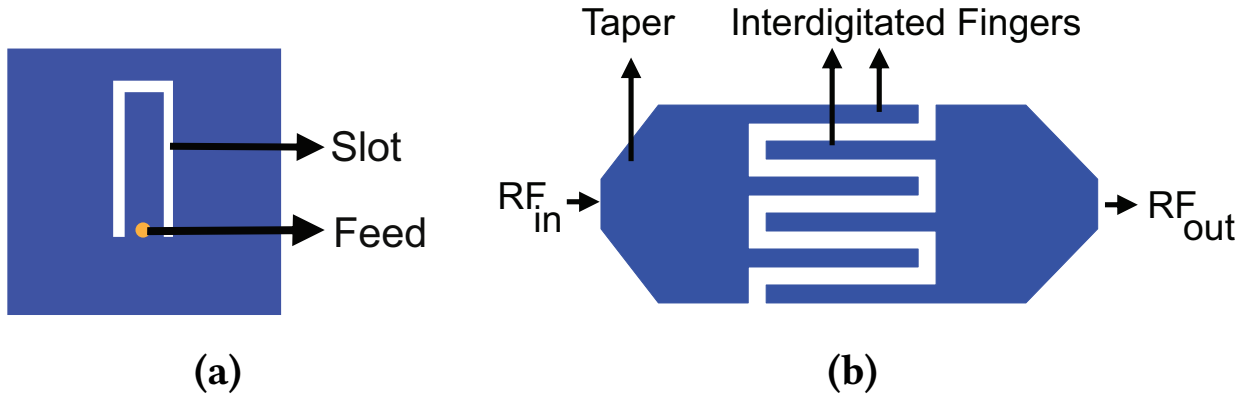

**Figure 5: Hardware Design:** (a) U Slot patch antenna, (b) Interdigitated capacitor.

*3.2.2* ***Re-generative Amplifier Design:*** As discussed earlier, the target is to create a stable amplifier with an input resistance as close to the $-ve$ characteristic impedance as possible in order to ensure that the circuit has very high gain while it doesn't oscillate.

The design process starts with choosing a suitable low-cost off-the-shelf transistor with a transition frequency in the higher mmWave region. In this work, we select the CEL CE3520K3 pHEMT, which operates well up to 26.5$GHz$ and naturally supports a common-source configuration. The S-parameter in the normal operating state of the transistor ($V_{DS}$ = 2V and $V_{GS}$ = -0.5V) in files provided by the manufacturer on their official website were imported into Keysight ADS simulation software, forming the basis for designing both the positive-feedback network and the impedance-matching circuitry.

The feedback circuit requires a footprint for an RF resistor used to tune the feedback factor of the amplifier and a capacitor needed to prevent a DC short between the input gate and output drain of the FET. At very high frequencies, discrete capacitors typically reach their self-resonance point, after which they begin to behave inductively. Although some recently developed capacitors maintain good performance in this frequency range, they are often costly and have poor quality factor.

To overcome these limitations, we design an interdigitated capacitor directly on the substrate [23]. Interdigitated capacitors consist of small finger-like structures with specific lengths and widths, separated by narrow gaps. In a simple parallel-plate capacitor, the capacitance $C = \frac{A\epsilon}{d}$ arises from the electric field stored between two plates of area $A$ separated by a dielectric of thickness $d$. Similarly, in an interdigitated capacitor, electric field vectors form between fingers connected to opposite polarities, producing the desired capacitive effect. The increased surface area of the interdigitated structure enables strong electric-field coupling and provides finer control over impedance. As you might expect, the resulting capacitance depends primarily on the finger length, finger width, number of fingers, and the dielectric properties of the substrate. More formally, the capacitance is given by

$$C = (\epsilon_r + 1)l(N - 3)(A_1 + A_2) \tag{3}$$

where $\epsilon_r$ is the relative permittivity of the dielectric, $l$ is the finger length, $N$ is the total number of fingers, and $A_1$ and $A_2$ are empirical parameters determined by the substrate height and finger width.

Another important point is that the structure also introduces some parasitic inductance due to the terminal transmission lines. This resulting LC network plays a crucial role in defining the frequency selectivity of our positive-feedback loop. The final structure is shown in Fig.5b. A resistor is placed in series with the capacitor, and the entire structure is connected through transmission lines to form the RLC network linking the drain and gate of the transistor.

The overall electrical length of the feedback path is chosen to be approximately one wavelength at the intended resonance frequency, ensuring a $0deg$ phase shift as discussed earlier. After implementing the feedback loop, an open stub is introduced at the gate input to assist with achieving proper input impedance matching. With the major components in place, the entire circuit is then carefully tuned—both in physical layout and in electrical parameters to accurately achieve resonance at the intended oscillation frequency.

Finally, some biasing lines with radial stubs are added to each side of the transistor to supply the constant $V_{DS}$ = 1.1V, on the one hand, and the $V_{GS}$—alternatively switched between 0V and -0.75V to create the FSK modulation—on the other.

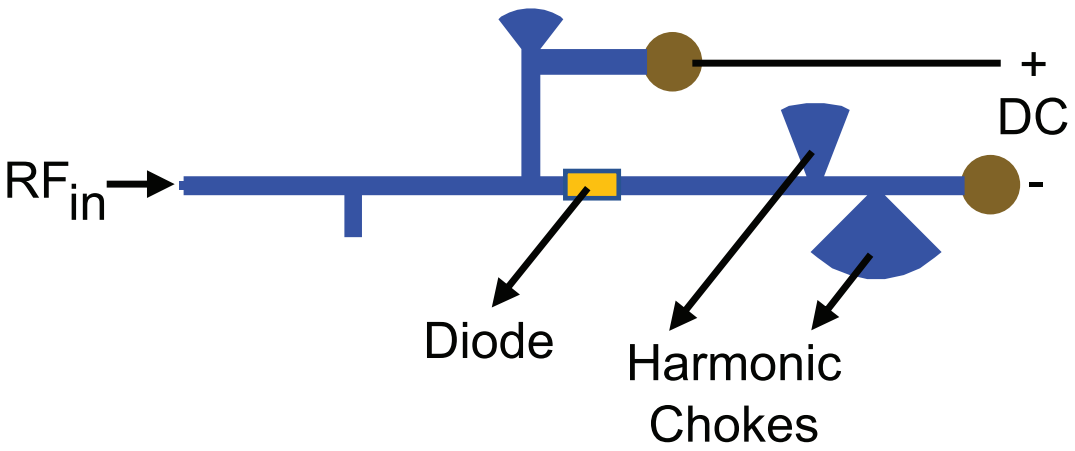

**Figure 6:** Designed passive rectifier schematic with Macom Schottky diode, harmonic stubs, and an L matching network.

*3.2.3* ***Rectifier Design:*** At mmWave frequencies, designing high-efficiency rectifiers presents significant challenges due to the limited availability of off-the-shelf rectifying components, particularly diodes, capable of operating at such high frequencies. To achieve optimal performance, these diodes must exhibit several key characteristics: low turn-on voltages, high cutoff frequencies, and minimal series resistances. In this design, the MA4E1317 diodes from Macom were selected. These diodes are gallium arsenide (GaAs) flip-chip Schottky barrier diodes, which are specifically designed to handle high-frequency applications. They feature a forward voltage of $0.7V$ and a relatively low series resistance of 4Ω, making them suitable for mmWave rectification.

The rectifier circuit incorporates an L-matching network, which is linked to a $\lambda/4$ extended radial stub. This stub serves to isolate the RF signal from the DC interrogation point. The cathode, in contrast, is connected to two additional radial stubs. These stubs are specifically designed to block higher-order harmonics [24]. By effectively filtering out unwanted harmonic frequencies, the stubs help enhance the overall efficiency of the rectifier circuit.

The circuit was simulated using Keysight ADS. A thinner (6 mils) Rogers 4350B substrate was used. Generally, thinner substrates lead to improved efficiencies, as they reduce dielectric losses and enable narrower microstrip line widths, allowing more compact routing. The rectifier design is shown in Fig.6.

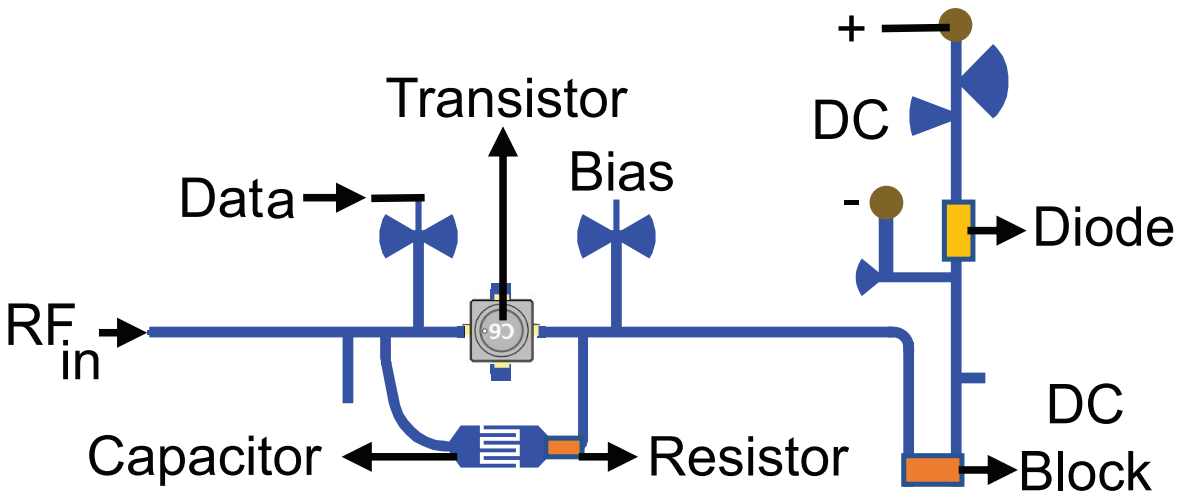

**Figure 7: Re-generative Rectifier:** Designed positive feedback amplifier-based rectifier schematic.

*3.2.4* ***Re-generative Receiver Design:*** As we saw earlier in Section 3.2.3, the sensitivities of passive rectifiers built using off-the-shelf diodes are limited to around −3 dBm. To put this into perspective, for a 26 GHz system with a standard 30 dBm EIRP transmitter and a 10 dB gain receiver antenna, the maximum achievable range would only be about 13 cm, even when evaluated using the ideal Friis transmission equation—which itself is strictly valid only in the far field and therefore optimistic in this scenario. Such a short operating distance is clearly insufficient for practical IoT applications, which typically require communication ranges on the order of several tens or hundreds of meters.

Unfortunately, the situation is not significantly better with commercially available active rectifiers. For instance, the widely used design in [25] exhibits a sensitivity of approximately −30 dBm, improving the range to roughly 3 m, but still falling short of many real-world deployment requirements. To push the sensitivity further, some form of active circuitry—such as an amplifier—becomes necessary. However, conventional mmWave amplifiers are often prohibitively expensive and power-hungry. To overcome this limitation, we integrate a regenerative amplifier with the rectifier, enabling substantial sensitivity enhancement while consuming relatively low power.

This integration, however, requires careful impedance planning, as regenerative amplifiers are highly sensitive to even small variations in their loading conditions. As a result, the design process must jointly optimize the amplifier while accounting for the rectifier's input impedance. Even with careful impedance matching and repeated optimization, the final performance is still influenced by fabrication tolerances and the limited accuracy of standard electromagnetic simulation tools, which often fail to capture the subtle geometric irregularities introduced during manufacturing that are especially critical in regenerative circuits due to their high sensitivity. To address these issues, the design was iteratively refined across multiple prototype–measurement cycles until it operated reliably at the desired frequency. Furthermore, to ensure stable operation, the final design includes a DC-block capacitor between the amplifier and rectifier, preventing unwanted DC leakage and maintaining proper isolation between the two circuits. The final integrated regenerative rectifier is shown in Fig. 7.

# 4 Components Characterization

## 4.1 Antenna

The U slot patch antenna was prototyped and its input reflection coefficient S11 (shown in Fig. 8) was measured, demonstrating wideband matching between 23.5 GHz and 27.2 GHz, while the antenna had a simulated gain of around $5dB$ in the entire impedance bandwidth range.

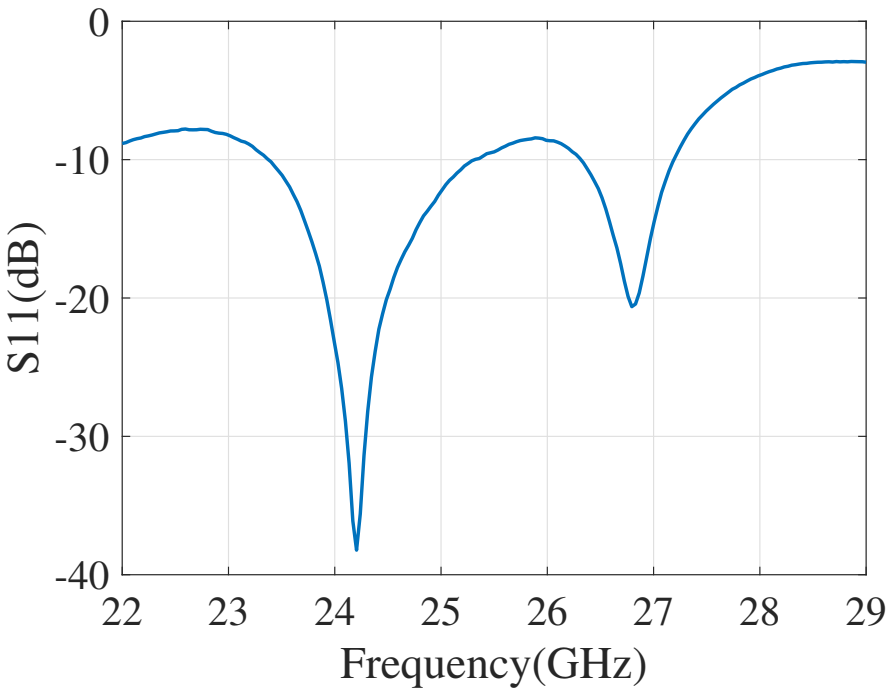

**Figure 8: Measured Antenna S11:** U slot patch antenna measured S11 demonstrating matching over the desired wide frequency range.

## 4.2 Regenerative Amplifier

The regenerative amplifier—fabricated on 6 mil Rogers 4350, using a 50Ω, 0402 resistor as its feedback resistance, and biased at $V_{GS} = 0V$—was characterized as a function of its input power level and drain voltage. The results, presented in Fig. 9, show that the amplifier can display a gain of more than 30 dB at input power levels of -40 dBm, before the saturation of its output reduces the effective gain down to 15 dB at input levels above -10 dBm. Although linearization techniques—such as predistortion—could improve the compression behavior, they add unnecessary complexity for our application, where the received RF input powers are typically very low. Furthermore, with an input power of -30$dBm$, the nominal operating voltage ($V_{DS}$ = 1.1V) biases the transistor in the saturation region, enabling high gain, while lower supply voltages shift operation toward the triode region, reducing amplification. The amplifier consumes a current of $7mA$ while being constantly modulated by FSK on its gate terminal.

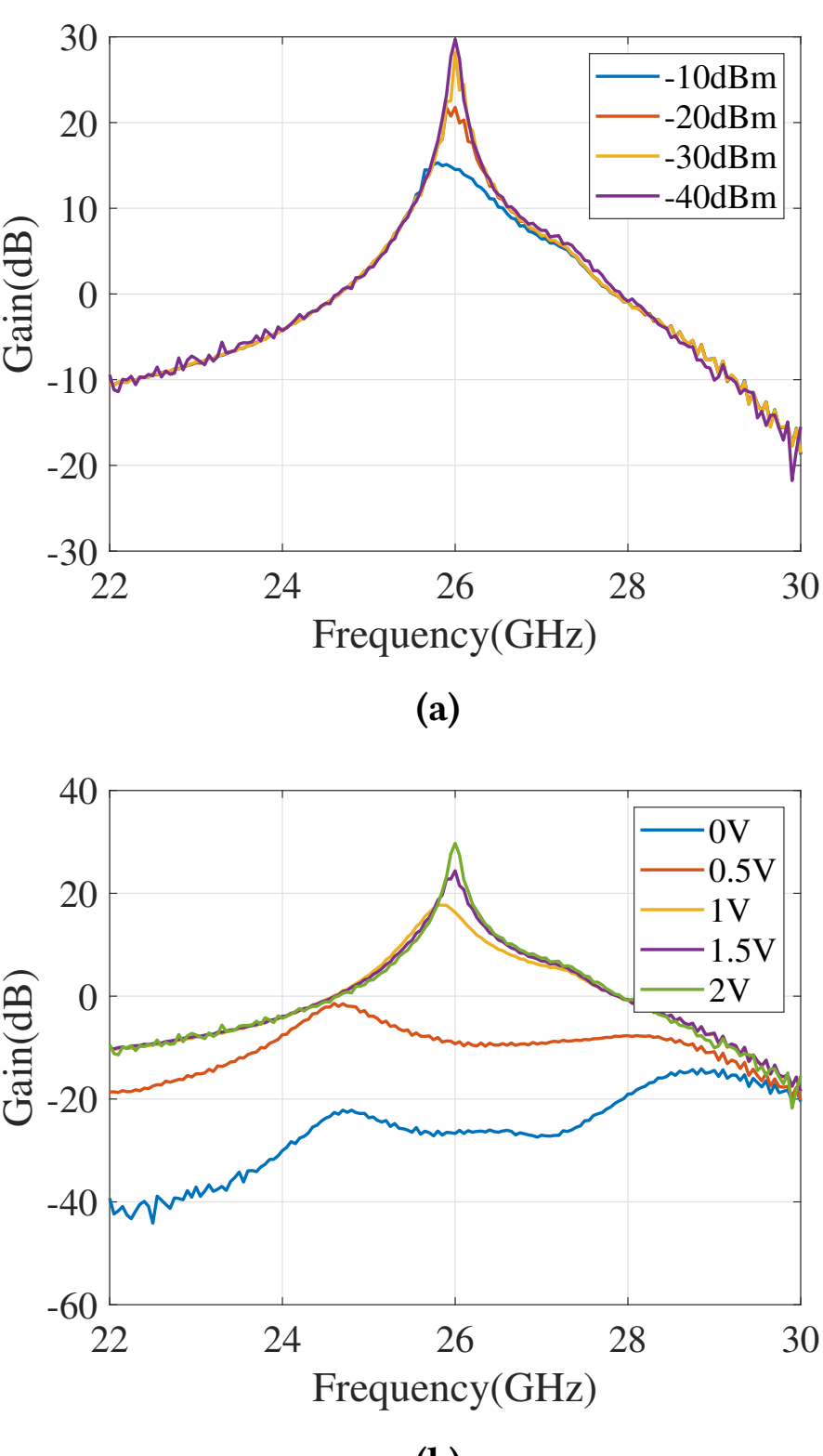

**Figure 9: Re-generative Amplifier Gain Performance under** (a) different input power levels, (b) different supply voltages ($V_{DS}$).

**High-Q regenerative amplifier and its full-duplex implications:** One of the key characteristics of the designed regenerative amplifier is its very high quality factor (Q). The Q factor of the amplifier, measured at a drain–source supply voltage of 2V and at RF input power levels below −35dBm, is nearly 210. A high Q factor means the circuit's performance degrades sharply with even small deviations in frequency. This property allows us to allocate uplink and downlink frequencies extremely close to one another without significant mutual interference, enabling more efficient spectrum usage—particularly valuable when multiple tags need to communicate simultaneously.

### 4.3 Rectifier and Regenerative Receiver

The open-circuited output voltage of the rectifier, measured as a function of input RF power, is shown in Fig. 10. The results demonstrate the effectiveness of the design in converting mmWave power into DC. As evident from Fig. 10, the output voltage increases with increasing input RF power; however, the rectifier's sensitivity remains around −3 dBm, which is insufficient for long-range deployment scenarios. To overcome this limitation and significantly enhance sensitivity, a regenerative amplifier is placed in front of the rectifier. The performance of this amplifier is presented in Sec. 4.2, with additional fine-tuning of both the amplifier and rectifier designs to ensure that the resonance frequency of the regenerative amplifier aligns closely with the frequency of peak rectifier efficiency.

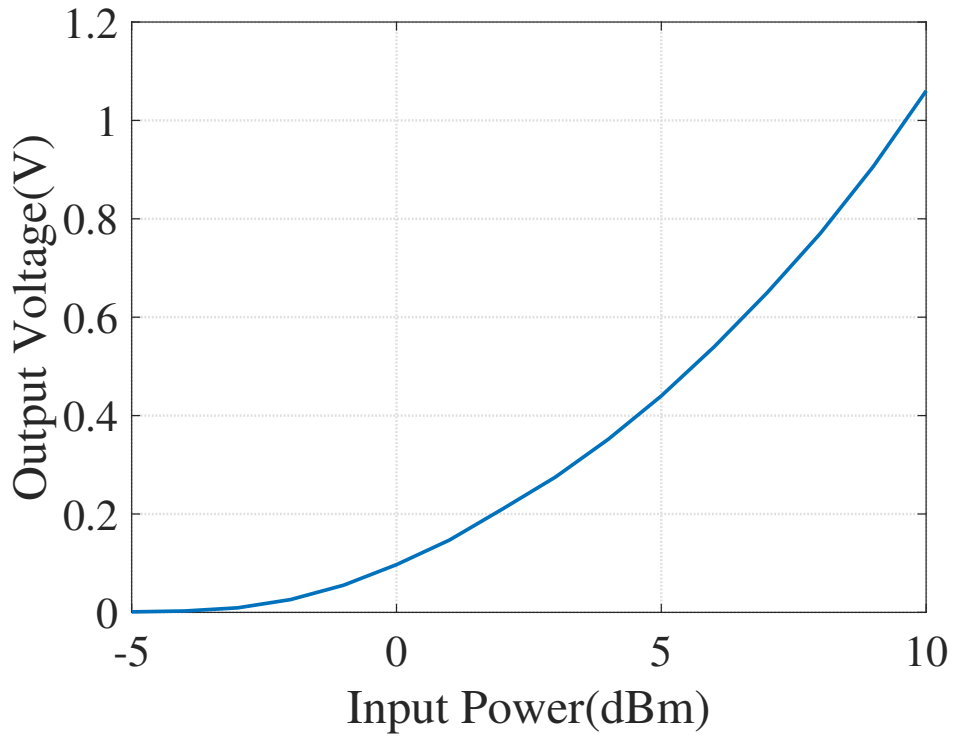

**Figure 10: Single Diode Passive Rectifier Performance:** Measured DC output voltages with different mmWave input power levels.

The finely tuned regenerative rectifier achieves an impressive sensitivity of approximately −60 dBm, representing a nearly 60 dB improvement over traditional passive rectifiers and a 30 dB gain over commercial active rectifiers. It should be noted that, while this may seem inconsistent with the gains shown in Fig. 9, the highly non-linear nature of the system can yield such behaviors. This enhancement finally enables operation over tens of meters, opening the door to practical long-range rectifier-based receivers.

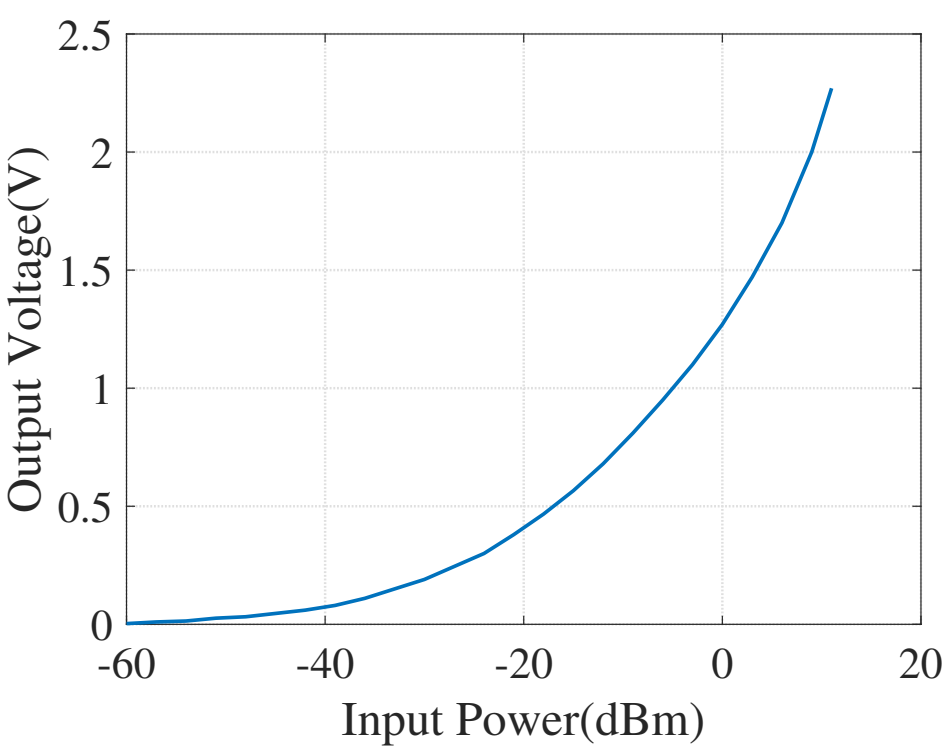

**Figure 11: Re-generative Rectifier Performance:** Measured DC output voltages vs significantly lower mmWave input power levels.

## 5 Implementation and Evaluation

### 5.1 Full-Duplex Tag

The full duplex tag, displayed in Fig.13, houses a $2 \times 2$ broadband patch antenna array, where the broadband patches are identical to the ones described earlier but combined into an array using Wilkinson power dividers. The complete system was first simulated and then fabricated on a four-layer Rogers 4350B board, using a 20 mil substrate on one side and a 6 mil substrate on the other, with all regenerative circuits included. A baseband circuit was also integrated. It includes the LD59150 low-dropout linear regulator, which regulates the battery-supplied power fed to the drains of the circuits. An oscillator was added to generate the modulation signals for uplink communication. The uplink baseband section also provides space to attach a COTS resistive sensor such as a photodiode, enabling analog sensing without any modifications to the tag.

### 5.2 Reader

The Armstrong tag was evaluated using the EVK2001 Evaluation Kit. A 40 MHz ASK-modulated baseband signal for downlink communication was generated by a standard arbitrary waveform generator. This baseband waveform was then upconverted by the EVK to a mmWave carrier in the 26.1 GHz–26.3 GHz range. The EVK's maximum EIRP of 35 dBm was used for all experiments. For uplink communication, the RFSoC 4×2 generated a simple sine-wave baseband excitation signal, which was similarly upconverted in the EVK to the mmWave band. The uplink backscattered data was recorded using the Analog Discovery 3 [26], a low-cost oscilloscope equipped with a 14-bit ADC supporting sampling rates up to 125 MSPS.

### 5.3 Data Processing

**Uplink:** The FSK data received at the EVK was sampled and processed using the Goertzel algorithm, an efficient method for computing individual DFT bins. Compared to an FFT, which computes the spectrum across all frequencies, the Goertzel algorithm evaluates the DFT only at specific frequencies of interest. Since FSK uses known, fixed tones to encode data, this approach is particularly well suited. After computing the magnitudes at the two FSK tones, the bit value is determined by selecting the frequency with the higher magnitude. Because no preamble or symbol boundary markers were used, a brute-force timing synchronization method was employed to identify the correct sampling offset.
**Downlink:** The envelope of the ASK signal is readily extracted by our regenerative rectifier, after which it is sampled and processed using standard ASK demodulation techniques. The first step is to synchronize the data to a clock signal operating at the same frequency as the ASK keying rate. Initial synchronization is performed by detecting zero crossings in the envelope, and then refined using a more robust approach: cross-correlating the preamble with the clock waveform. The resulting phase offset is applied to the clock before decoding. Decoding is performed by applying a fixed threshold to the low-pass filtered envelope and sampling the amplitude at each clock pulse. If the sampled value exceeds the threshold, the bit is decoded as a 1; otherwise, it is decoded as a 0.

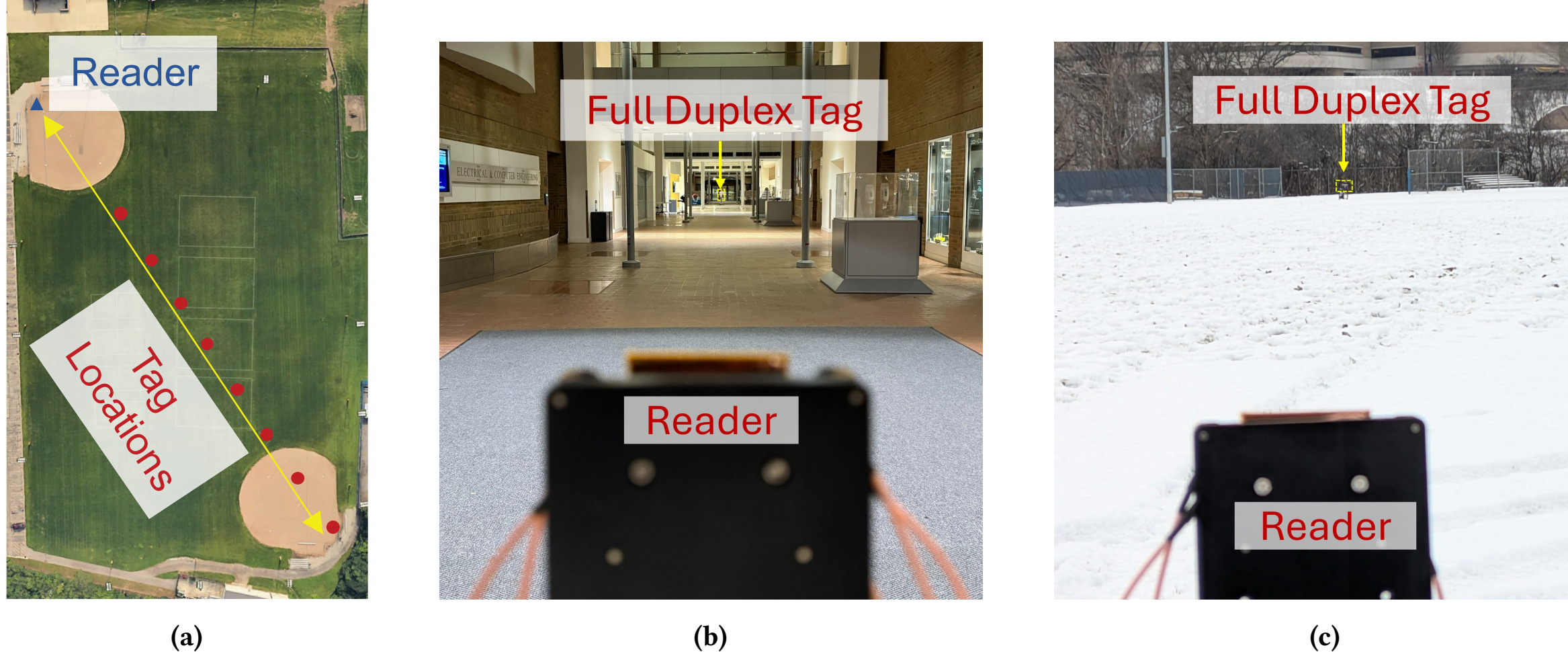

(a) (b) (c)

**Figure 12:** (a) Field map of the outdoor experimental setup, (b) indoor setup, (c) outdoor setup.

## 5.4 Evaluation

**Full-Duplex Indoor Tests:** We evaluated the system in a 90$m$-long atrium, in an environment rich in multipath. In all experiments, the tag and reader were placed facing each other, as shown in Fig.12b. The tests were performed at different distances from 5$m$ to all the way up to the entire length of the atrium. In all these tests, the tag was operated in full duplex mode—both transmitting and receiving front-ends were being powered and modulated.

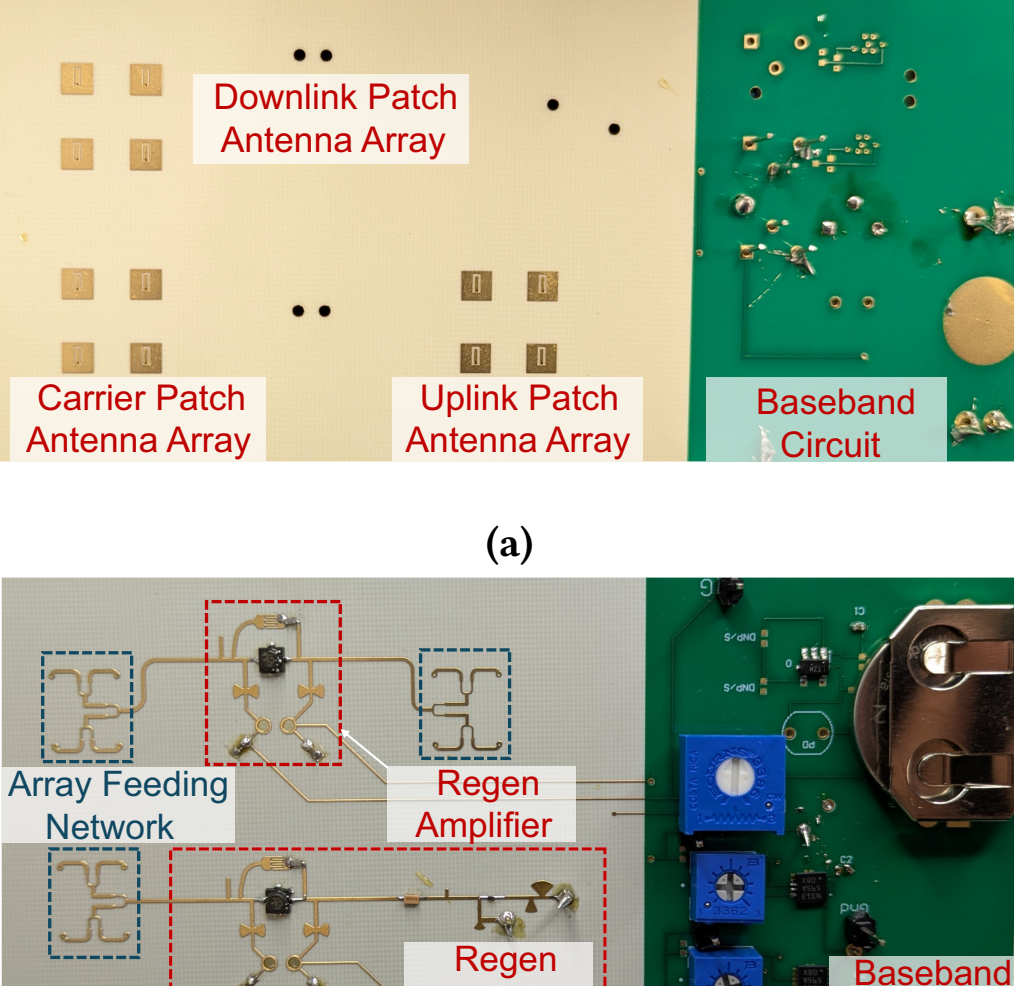

(a)

(b)

**Figure 13: Armstrong Tag:** (a) photo of the top layer featuring the designed patch antenna arrayed in a 2x2 configuration, and used for the carrier, uplink, and downlink, (b) photo of the bottom layer featuring the antenna array feeding network, the regenerative amplifier on the uplink, the regenerative rectifier on the downlink, biasing circuitry and a coin cell battery.

**Outdoor Downlink Tests:** The outdoor environment consisted of a 200m long (diagonally) open field, whose satellite image is shown in Fig. 12a. On the day of the experiments, the ambient temperature ranged from 0°C to 7°C. Line of sight between the reader and the tag was consistently maintained throughout the experiments. Both the reader and the tag were mounted approximately 1.3m above the ground. While the reader remained stationary, the tag was moved to different locations, as indicated in Fig. 12a. It is important to note that the ground was not exactly even and this may also have an effect on the link performance. Only the downlink circuitry of the tag was powered during these measurements, due to the testing ranges exceeding the maximum range of the uplink.

## 6 Results

Different sets of experiments were conducted to assess the SNR and BER of the bidirectional communication links at varying ranges and data rates in both indoor and outdoor settings. The tag's frequency sensitivity was also evaluated to help us understand the frequency isolation necessary between uplink and downlink for full-duplex operation and quantify the frequency multiplexing density allowed by the architecture for dense deployments. Furthermore, the system's performance in NLoS and dynamic settings was evaluated.

### 6.1 SNR versus Bit Rate and Range

First, the SNR of the wireless links was measured at ranges up to 88m indoors and 200m outdoors for data rates ranging from 20 to 60 Kbps in downlink and 1 to 20 Kbps in uplink. The results of this experiment are shown in Fig. 14a and Fig. 15a. We make the following remarks:

- In downlink, the SNR unsurprisingly approximately obeys the one-way Friss power law, with its magnitude decreasing approximately 6 dB for every doubling of the range. Likewise, the SNR at 60 Kbps is on average about 4 dB lower than at 20 Kbps, where 4.8 dB would be expected. Even at 200m, the SNR for 60 Kbps communications maintains an acceptable value of 17 dB, thereby demonstrating that the impressive 200m range achieved could further be extended.
- In uplink, the SNR more interestingly also obeys the one-way Friss law despite traveling back and forth from and to the reader. This can be explained due to the saturating effect of the regenerative amplifier: the amplifier locks onto the carrier and amplifies the received signal until reaching a steady and constant power level, regardless of the reception range, thereby effectively behaving like an active transmitter.

### 6.2 BER across Bit Rate and Range

Next, the bit error rates (BER) of the system operating at different bit rates and at different ranges were experimentally assessed. Here too, the experiments were conducted indoors until 88m, followed by outdoor measurements up to 200m. The results of this experiment are shown in Fig. 14b and Fig. 15b. We make the following remarks:

- In downlink, most BERs remain under $10^{-4}$ for all modulation rates until 35m, while 1 Kbps remains under this threshold until 75m in range. These quickly rise thereafter, before reaching $10^{-1}$ at 200m.
- In uplink, all BERs remain under $10^{-4}$ beyond a range of 45m, while 1 Kbps allows under $10^{-1}$ even up to the maximum range of 88m.

### 6.3 Frequency Sensitivity and Multi-Tag Isolation

**Full-duplex isolation:** In any full-duplex system, it is important to minimize the potential interference between the uplink and downlink, as simultaneous transmission and reception can degrade signal quality. In the Armstrong tag tested to produce Fig. 14 and Fig. 15, downlink and uplink resonances were set at 26.35GHz and 26.41GHz, respectively. In all the testing that was conducted, no visible difference in the SNRs or BERs were observed in the 88m of range where both up and down links could simultaneously operate. It is also critical—especially in the downlink, where frequency multiplexing of the ASK data cannot readily and natively be achieved by the rectifier—to assess the ability for multiple tags to be multiplexed for communications.

**Multi-tag isolation:** In Fig. 14c and Fig. 15c are shown the simultaneous SNRs received by two tags whose regenerative front ends were tuned at slightly different frequencies in downlink and uplink, respectively. This experiment, conducted exclusively indoors at a distance of $5m$ with a lower transmit EIRP, varied the carrier frequency used by the reader in either the uplink or downlink and measured its effect on the SNRs received from and by both tags. We make the following remarks on the results of these experiments:

- In downlink, the SNR quickly decreases by more than 10 dB over 100 MHz of carrier offset. This is expected, as the voltage produced by the rectifier decreases exponentially with the reduction in input power level, exacerbated by the de-tuning of the amplifier.
- In uplink, the decrease is even sharper, with the SNR decreasing by 10 dB within 10 MHz. This can be attributed to the increasingly tenuous coherence between the reader's carrier and the modulated signal re-emitted by the tag, as the offset between its and the amplifier's frequencies rises. item The two tags appear to behave linearly independently—the behavior of their proximal combination is the sum of their individual responses—i.e. they show no sign of affecting each other.

### 6.4 SNR-BER in NLoS Conditions

The performance of Armstrong was evaluated under NLoS conditions in both downlink and uplink scenarios at a range of 5m. The same experimental setup used for the aforementioned LoS measurements was employed, with obstacles introduced 50cm in front of the reader. These obstacles included a 2mm thick glass frame, a 3cm-thick dry wood panel, an absorbing foam material [27], a person, and a 20cm-thick concrete wall, as shown in Fig. 16a. Figures 16c and 16b present the relative SNR and BER results for both downlink and uplink across the five NLoS scenarios, compared to the LoS case. As is commonly observed from mmWave systems, Armstrong achieves good penetration through untempered glass and wood, but suffers much more serious attenuation through concrete walls or high water content bodies. This behavior—along with the system's robustness to multipath—demonstrates its ability to maintain a relatively deterministic link in even cluttered environments, allowing its deployments to be engineered and dimensioned to provide the coverage required by each application.

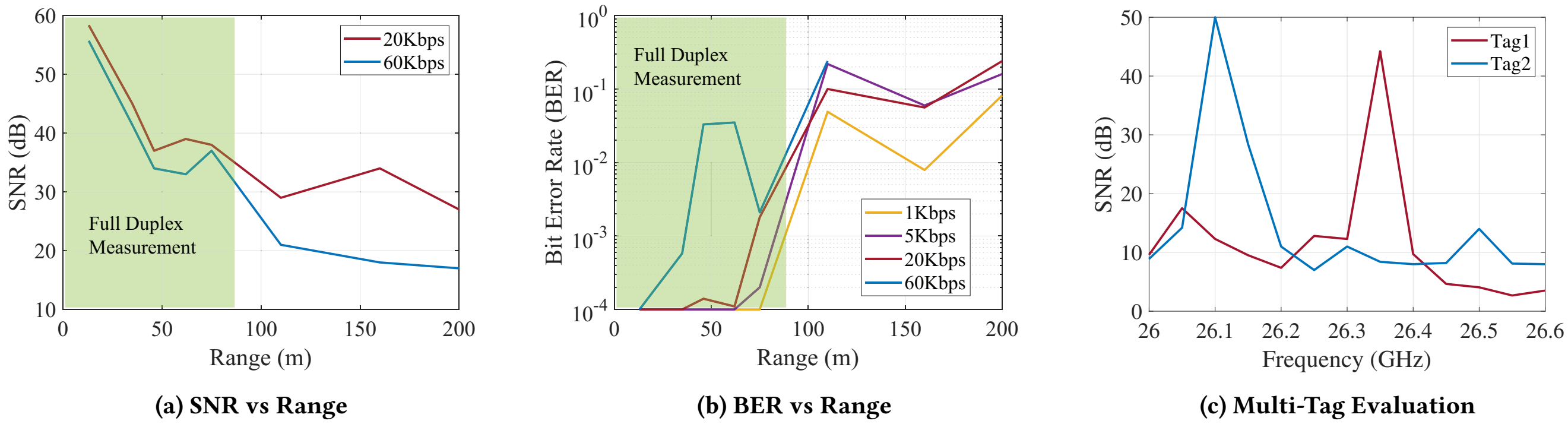

**Figure 14: Downlink Performance of Armstrong:** (a) & (b) The tag demonstrates reliable indoor and outdoor operation up to distances of 200 m. (c) The tags exhibit high frequency selectivity, allowing their frequency multiplexing within less than 100 MHz.

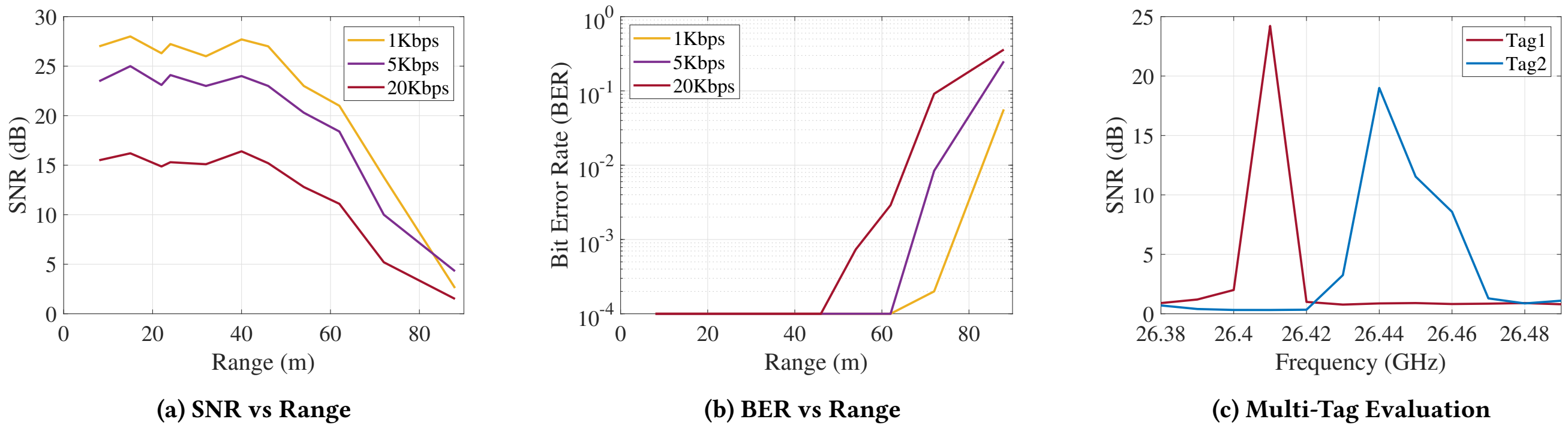

**Figure 15: Uplink Performance of Armstrong:** (a) & (b) The tag demonstrates efficient backscattering of FSK-modulated data over long indoor ranges. (c) The tags exhibit high frequency selectivity, allowing their frequency multiplexing within less than 10 MHz.

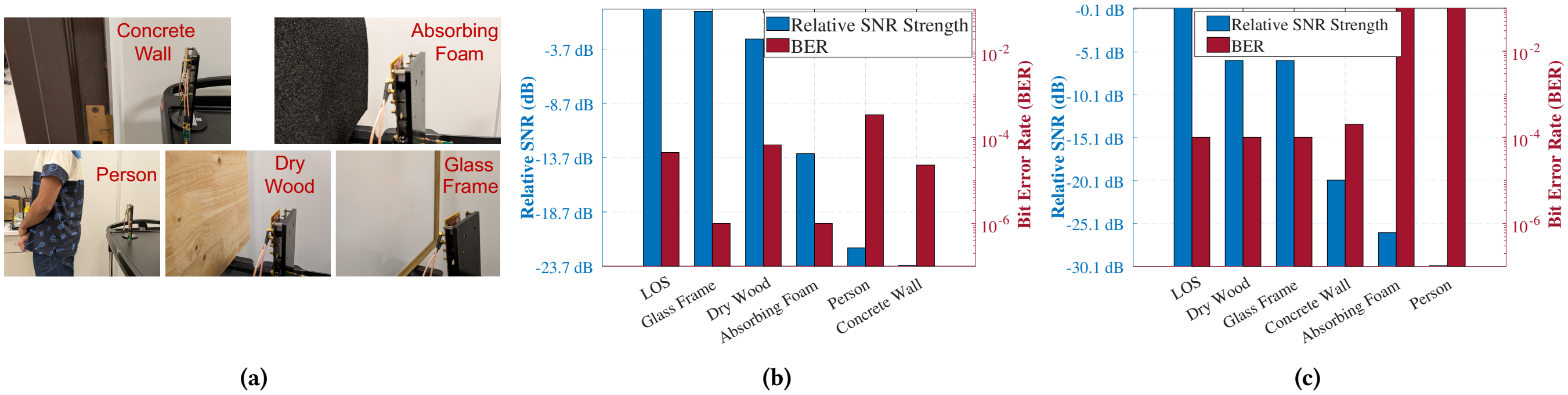

**Figure 16: Armstrong's Downlink and Uplink Performance in NLoS Conditions** (a) with different obstacles, (b) relative SNR and BER results during downlink, (c) relative SNR and BER results during uplink.

## 6.5 SNR-BER in Mobile Settings

The performance of Armstrong was also characterized in mobile settings for both downlink and uplink scenarios at a range of approximately 2m. This was achieved by mounting the tag on a toy train, as shown in Fig. 17a, and varying its velocity from 20cm/s to 75cm/s. All measurements were performed with the train moving radially towards the reader at the specified velocities. Figs. 17b and 17c present the corresponding SNR and BER results for both downlink and uplink as a function of velocity. While the links maintain a usable BER below $10^{-1}$ under these dynamic conditions, a noticeable degradation in communication performance is observed. Addressing this limitation and improving mobility robustness of the proposed communication schemes is an important direction for future work.

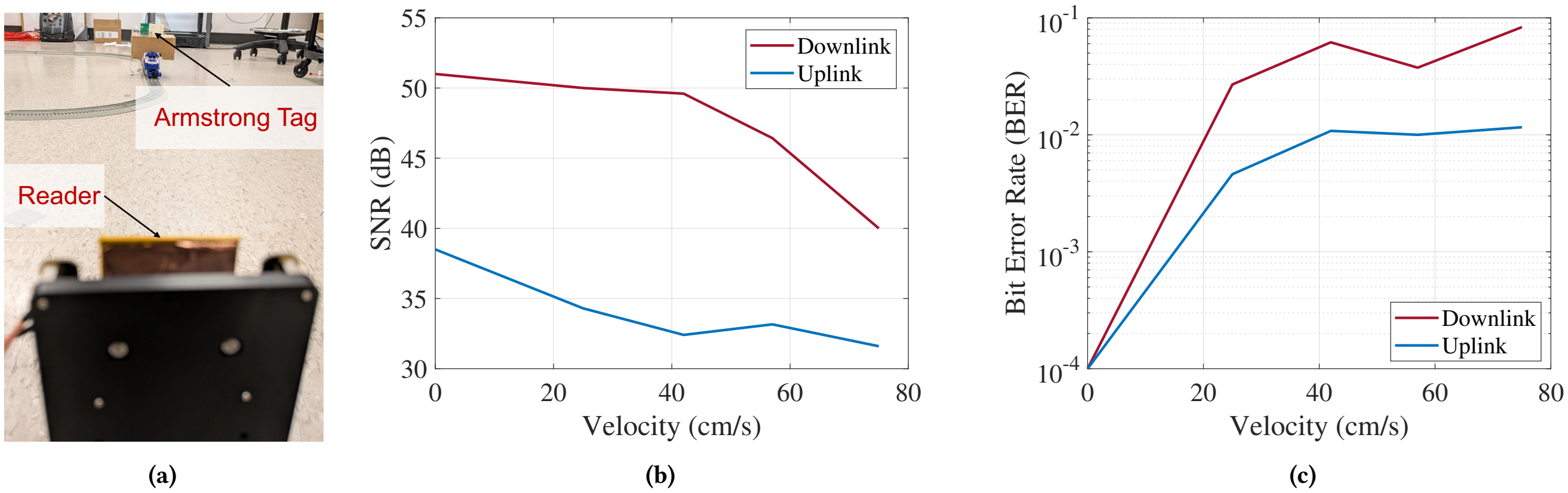

**Figure 17: Armstrong’s Downlink and Uplink Performance under Mobility:** (a) Armstrong is securely placed on a toy train and moved with different velocities from $25cm/s$ to $75cm/s$, (b) SNR vs velocity results, (c) BER vs velocity results.

| Systems | Tag Power Draw | Tag Cost (USD) | Reader EIRP (dBm) | Tag antenna gain (dBi) | Max Uplink Range and Data Rate | Max Downl. Range and Data Rate | Full Duplex Operation |
|---|---|---|---|---|---|---|---|
| **Armstrong** | **48.4$m$W** | 5 | **35** | **9** | **88m @20Kbps** | **200m @20Kbps** | **Yes** |
| **BiScatter [11]** | 135$m$W | 400 | 20* | 16 | 7$m$ @0.1Kbps* | 7$m$ @60Kbps | No |
| **MilBack [12]** | 270$m$W | 500 | 47 | 13 | 8$m$ @40Mbps | 10$m$ @36Mbps | No |

**Table 1: Comparison with the state-of-the-art systems. (* best estimates)**

## 7 Related Work

**RF Backscatter Systems:** There is a substantial body of work exploring backscatter tags operating at sub-6 GHz frequencies. Several efforts have examined the use of commercial RFID tags to enable one-way communication, either for localization [28–32] or for transmitting small amounts of identification data to support environmental sensing tasks [33]. However, these systems do not aim to support meaningful data transmission. Other works have proposed more advanced backscatter tags capable of handling higher-order modulations, such as QAM [34], but these efforts also focus exclusively on uplink communication. A particularly interesting and more relevant effort is the work by Liu et al. [35], which, for the first time, demonstrates a tag capable of instantaneous two-way communication. Their design uses a simple single-antenna architecture and enables ambient inter-tag communication using a shared RF source. However, this system targets low-throughput communication intended for small data exchanges, and because it is based on passive RFID tags, its operational range is inherently limited to just a few meters.

**mmWave Backscatter Systems:** Recently, several works such as [7, 8, 10, 16, 36] have introduced techniques for localizing backscatter tags using simple square-wave baseband modulation, without the need to transmit meaningful data. Researchers have also explored mmWave backscatter for achieving high throughput and extending the capabilities of RFID-style systems. For instance, mmTag [37] demonstrates Gbps-class backscatter links using custom tag and reader designs, proving the feasibility of high-rate mmWave uplink (tag-to-reader) communication. More recent efforts such as [11, 12] are closer to our work, as they support both uplink and downlink communication. MilBack [12] employs a frequency-selective leaky-wave antenna connected to a switch and an active envelope detector. For downlink communication, the reader must first determine the orientation of the tag and then select the appropriate frequencies to use, necessitating a handshake procedure. Additionally, FMCW waveforms are used by the reader to facilitate localization.

BiScatter [11] advances this idea by enabling downlink communication through chirp spread spectrum (CSS) modulation—similar to that used in LoRa—thereby reducing the reader's hardware complexity to a single FMCW radar. The BiScatter tag incorporates multiple delay lines, a switch, and an active envelope detector to assist in chirp slope detection, ultimately enabling two-way communication. However, these systems still operate in half-duplex mode, focusing on either uplink or downlink performance at relatively short ranges, but not supporting simultaneous bidirectional communication.

**Reflection and Regenerative Amplifiers for Backscatter:** Reflection and regenerative amplifiers have long been explored as a means to achieve reflection coefficients greater than unity and thereby improve backscatter performance. Prior work has shown that such amplifiers can significantly boost tag SNR and extend communication range. Broadly, two amplifier architectures are commonly used: one based on transistor implementations [38, 39], and the other based on tunnel diodes. Tunnel-diode backscatter systems exploit the negative-resistance region of the device, allowing the diode to function simultaneously as an oscillator and a reflection amplifier. When biased in this region, the diode produces a regenerative effect that enhances the backscattered signal, enabling longer-range communication with exceptionally low power consumption [40, 41]. For example, TunnelScatter [41] demonstrates a tunnel-diode oscillator capable of transmitting even without an ambient carrier, while also serving as a reflection amplifier when ambient signals are weak.

Despite their advantages, tunnel diodes are very costly and do not scale well to high mmWave frequencies. Recently, researchers have begun exploring regenerative amplifiers operating directly at mmWave frequencies. For instance, a 24 GHz super-regenerative amplifier has been proposed for tag front-ends [42]. However, these designs are still restricted to uplink-only communication and rely on custom ICs fabricated in specialized 135 nm BiCMOS processes, limiting their practicality for low-cost IoT backscatter tags.

**mmWave rectifiers:** mmWave rectifier literature can be broadly classified into two broad categories. One uses off the shelf diodes [43–46] for rectification while the other uses custom CMOS IC's[47, 48]. The rectifiers with off the shelf diodes have very low sensitivities of 0 dBm, since off the shelf diode have high turn on voltages. The sensitivities are improved to around -15 dBm with the lower thresholds obtained in CMOS technology along with some external cold starting techniques but they still fall short of the sensitivities required for long range communication. Nevertheless we believe we can incorporate such on chip rectifiers to further increase our sensitivities.

## 8 Limitations and Conclusion

This paper introduces Armstrong, an architecture enabling long-range full-duplex backscatter low-cost IoT tags for mmWave ISAC implementations. The system is demonstrated to achieve the first demonstration of full duplex mmWave backscatter as well as almost two orders of magnitude longer downlink ranges than the art at 100x less the cost, as shown in Table 1. However, the reported system suffers from certain limitations that motivate further efforts:

• **Tuning Sensitivity:** The regenerative amplifiers—like their close tunnel-diodes-based cousins—are very narrow band and somewhat influenced by environmental factors like temperature. Methods designed to account for these uncertainties will have to be developed.

• **Uplink Range:** Given the high gain provided by the amplifier, the uplink range demonstrated in this work is somewhat underwhelming. This is likely due to the signal quickly losing coherence with the reader's carrier as the amplifier oscillates. Its design will have to be adjusted to reduce this effect.

• **Localization:** While the system is motivated by its intended use in ISAC implementations, this work does not report a display of such a capability. Future work will focus on reporting the performance of this system in such a context, thereby providing a true ISAC demonstration. The implementation of base-station waveforms capable of simultaneously imaging passive targets could also be explored.

• **Mobility:** The tags would, furthermore, be required to offer all the required capabilities while mounted on mobile equipment and vehicles. As shown in Sec. 6.5, mobility can have detrimental effects on the communications of the proposed system. As the research evolves, the influence of mobility would have to be mitigated using more robust modulation and detection schemes.

While many important problems remain unexplored, Armstrong provides a new reference system architecture for the emergence of mmWave ISAC systems for IoT applications and blazes a trail for exciting new research landscapes.

## Acknowledgments

We thank our shepherd and the anonymous reviewers for their valuable and constructive feedback. We also thank Longyu Guo and Yunfei Liu from the Beam Dynamics group for their assistance with the experiments.

## References

[1] W. Saad, M. Bennis, and M. Chen, "A vision of 6g wireless systems: Applications, trends, technologies, and open research problems," *IEEE network*, vol. 34, no. 3, pp. 134–142, 2019.

[2] Y. Cui, F. Liu, X. Jing, and J. Mu, "Integrating sensing and communications for ubiquitous iot: Applications, trends, and challenges," *IEEE network*, vol. 35, no. 5, pp. 158–167, 2021.

[3] F. Liu, Y. Cui, C. Masouros, J. Xu, T. X. Han, Y. C. Eldar, and S. Buzzi,

"Integrated sensing and communications: Toward dual-functional wireless networks for 6g and beyond," *IEEE journal on selected areas in communications*, vol. 40, no. 6, pp. 1728–1767, 2022.

[4] J. Wang, N. Varshney, C. Gentile, S. Blandino, J. Chuang, and N. Golmie, "Integrated sensing and communication: Enabling techniques, applications, tools and data sets, standardization, and future directions," *IEEE Internet of Things Journal*, vol. 9, no. 23, pp. 23 416–23 440, 2022.

[5] 3GPP, "Study on NR positioning support," TR 38.855, Technical Report 16.0.0, Tech. Rep., 2019.

[6] R. Mundlamuri, "Localization and sensing in 5g nr and beyond," Ph.D. dissertation, Sorbonne Université, 2025.

[7] K. M. Bae, N. Ahn, Y. Chae, P. Pathak, S.-M. Sohn, and S. M. Kim, "Omniscatter: extreme sensitivity mmwave backscattering using commodity fmcw radar," in *Proceedings of the 20th Annual International Conference on Mobile Systems, Applications and Services*, ser. MobiSys '22. New York, NY, USA: Association for Computing Machinery, 2022, p. 316–329. [Online]. Available: https://doi.org/10.1145/3498361.3538924

[8] E. Soltanaghaei, A. Prabhakara, A. Balanuta, M. Anderson, J. M. Rabaey, S. Kumar, and A. Rowe, "Millimetro: mmwave retro-reflective tags for accurate, long range localization," in *Proceedings of the 27th Annual International Conference on Mobile Computing and Networking*, ser. MobiCom '21. New York, NY, USA: Association for Computing Machinery, 2021, p. 69–82. [Online]. Available: https://doi.org/10.1145/3447993.3448627

[9] S. Harisha, J. G. Hester, and A. Eid, "Long-range mmid localization and orientation sensing via frequency-divided beam multiplexing," in *2024 IEEE International Conference on RFID (RFID)*, 2024, pp. 1–6.

[10] S. Harisha, J. G. D. Hester, and A. Eid, "Dragonfly: Single mmwave radar 3d localization of highly dynamic tags in gps-denied environments," in *Proceedings of the 31st Annual International Conference on Mobile Computing and Networking*, ser. ACM MOBICOM '25. New York, NY, USA: Association for Computing Machinery, 2025, p. 1136–1150. [Online]. Available: https://doi.org/10.1145/3680207.3765269

[11] R. Okubo, L. Jacobs, J. Wang, S. Bowers, and E. Soltanaghai, "Integrated two-way radar backscatter communication and sensing with low-power iot tags," in *Proceedings of the ACM SIGCOMM 2024 Conference*, ser. ACM SIGCOMM '24. New York, NY, USA: Association for Computing Machinery, 2024, p. 327–339. [Online]. Available: https://doi.org/10.1145/3651890.3672226

[12] H. Lu, M. Mazaheri, R. Rezvani, and O. Abari, "A millimeter wave backscatter network for two-way communication and localization," in *Proceedings of the ACM SIGCOMM 2023 Conference*, ser. ACM SIGCOMM '23. New York, NY, USA: Association for Computing Machinery, 2023, p. 49–61. [Online]. Available: https://doi.org/10.1145/3603269.3604873

[13] E. Sharp and M. Diab, "Van atta reflector array," *IRE Transactions on Antennas and Propagation*, vol. 8, no. 4, pp. 436–438, 1960.

[14] W. Rotman and R. Turner, "Wide-angle microwave lens for line source applications," *IEEE Transactions on Antennas and Propagation*, vol. 11, no. 6, pp. 623–632, 1963.

[15] A. Eid, J. G. D. Hester, and M. M. Tentzeris, "Rotman lens-based wide angular coverage and high-gain semipassive architecture for ultralong range mm-wave rfids," *IEEE Antennas and Wireless Propagation Letters*, vol. 19, no. 11, pp. 1943–1947, 2020.

[16] K. M. Bae, H. Moon, S.-M. Sohn, and S. M. Kim, "Hawkeye: Hectometer-range subcentimeter localization for large-scale mmwave backscatter," in *Proceedings of the 21st Annual International Conference on Mobile Systems, Applications and Services*, 2023, pp. 303–316.

[17] S. J. Douglas, *Inventing american broadcasting, 1899-1922*. Johns Hopkins University Press, 1987.

[18] EVK02001 RF Module EVK, https://www.sivers-semiconductors.com/wireless/evk02001/.

[19] RFSoC 4x2 Kit , https://www.amd.com/en/corporate/university-program/aup-boards/rfsoc4x2.html.

[20] K. F. Lee, S. L. S. Yang, A. A. Kishk, and K. M. Luk, "The versatile u-slot patch antenna," *IEEE Antennas and Propagation Magazine*, vol. 52, no. 1, pp. 71–88, 2010.

[21] K. Y. Lam, K.-M. Luk, K. F. Lee, H. Wong, and K. B. Ng, "Small circularly polarized u-slot wideband patch antenna," *IEEE Antennas and Wireless Propagation Letters*, vol. 10, pp. 87–90, 2011.

[22] H. Wang, X. Huang, and D. Fang, "A single layer wideband u-slot microstrip patch antenna array," *IEEE antennas and wireless propagation letters*, vol. 7, pp. 9–12, 2008.

[23] G. Alley, "Interdigital capacitors and their application to lumped-element microwave integrated circuits," *IEEE Transactions on Microwave Theory and Techniques*, vol. 18, no. 12, pp. 1028–1033, 1970.

[24] S. Ladan and K. Wu, "Nonlinear modeling and harmonic recycling of millimeter-wave rectifier circuit," *IEEE Transactions on Microwave Theory and Techniques*, vol. 63, no. 3, pp. 937–944, 2015.

[25] EVAL-ADL6010 , https://www.analog.com/en/resources/evaluation-hardware-and-software/evaluation-boards-kits/eval-adl6010.html.

[26] Analog Discovery 3 , https://files.digilent.com/datasheets/Analog-Discovery-3-Datasheet.pdf.

[27] Microwave Absorbing Foam, https://www.laird.com/products/absorbers/microwave-absorbing-foams/single-layer-foams/eccosorb-hr.

[28] L. Yao, W. Ruan, Q. Z. Sheng, X. Li, and N. J. Falkner, "Exploring tag-free rfid-based passive localization and tracking via learning-based probabilistic approaches," in *Proceedings of the 23rd ACM International Conference on Conference on Information and Knowledge Management*, ser. CIKM '14. New York, NY, USA: Association for Computing Machinery, 2014, p. 1799–1802. [Online]. Available: https://doi.org/10.1145/2661829.2661873

[29] J. Wang and D. Katabi, "Dude, where's my card? rfid positioning that works with multipath and non-line of sight," in *Proceedings of the ACM SIGCOMM 2013 Conference on SIGCOMM*, ser. SIGCOMM '13. New York, NY, USA: Association for Computing Machinery, 2013, p. 51–62. [Online]. Available: https://doi.org/10.1145/2486001.2486029

[30] R. Nandakumar, V. Iyer, and S. Gollakota, "3d localization for sub-centimeter sized devices," in *Proceedings of the 16th ACM Conference on Embedded Networked Sensor Systems*, ser. SenSys '18. New York, NY, USA: Association for Computing Machinery, 2018, p. 108–119. [Online]. Available: https://doi.org/10.1145/3274783.3274851

[31] J. Wang, D. Vasisht, and D. Katabi, "Rf-idraw: virtual touch screen in the air using rf signals," in *Proceedings of the 2014 ACM Conference on SIGCOMM*, ser. SIGCOMM '14. New York, NY, USA: Association for Computing Machinery, 2014, p. 235–246. [Online]. Available: https://doi.org/10.1145/2619239.2626330

[32] Y. Ma, N. Selby, and F. Adib, "Minding the billions: Ultra-wideband localization for deployed rfid tags," in *Proceedings of the 23rd Annual International Conference on Mobile Computing and Networking*, ser. MobiCom '17. New York, NY, USA: Association for Computing Machinery, 2017, p. 248–260. [Online]. Available: https://doi.org/10.1145/3117811.3117833

[33] Y.-H. Su, J. Ren, Z. Qian, D. Fouhey, and A. Sample, "Tomoid: A scalable approach to device free indoor localization via rfid tomography," in *IEEE INFOCOM 2023 - IEEE Conference on Computer Communications*, 2023, pp. 1–10.

[34] J. Zhao, G. Wang, D. Li, S. Xu, X. Guo, and Y. Li, "Optimal 4qam backscatter modulation for passive uhf crfid tags," *Physical Communication*, vol. 66, p. 102421, 2024. [Online]. Available: https://www.sciencedirect.com/science/article/pii/S1874490724001393

[35] V. Liu, V. Talla, and S. Gollakota, "Enabling instantaneous feedback

with full-duplex backscatter," in *Proceedings of the 20th Annual International Conference on Mobile Computing and Networking*, ser. MobiCom '14. New York, NY, USA: Association for Computing Machinery, 2014, p. 67–78. [Online]. Available: https://doi.org/10.1145/2639108.2639136

[36] J. G. D. Hester and M. M. Tentzeris, "Inkjet-printed flexible mm-wave van-atta reflectarrays: A solution for ultralong-range dense multitag and multisensing chipless rfid implementations for iot smart skins," *IEEE Transactions on Microwave Theory and Techniques*, vol. 64, no. 12, pp. 4763–4773, 2016.

[37] M. H. Mazaheri, A. Chen, and O. Abari, "mmtag: a millimeter wave backscatter network," in *Proceedings of the 2021 ACM SIGCOMM 2021 Conference*, ser. SIGCOMM '21. New York, NY, USA: Association for Computing Machinery, 2021, p. 463–474. [Online]. Available: https://doi.org/10.1145/3452296.3472917

[38] J. Kimionis, A. Georgiadis, A. Collado, and M. M. Tentzeris, "Inkjet-printed reflection amplifier for increased-range backscatter radio," in *44th European Microwave Conference (EuMC), 2014*, 2014, pp. 53–56.

[39] ——, "Enhancement of rf tag backscatter efficiency with low-power reflection amplifiers," *IEEE Transactions on Microwave Theory and Techniques*, vol. 62, no. 12, pp. 3562–3571, 2014.

[40] F. Amato and G. D. Durgin, "Tunnel diodes for backscattering communications," in *2018 2nd URSI Atlantic Radio Science Meeting (AT-RASC)*, 2018, pp. 1–3.

[41] A. Varshney, A. Soleiman, and T. Voigt, "Tunnelscatter: Low power communication for sensor tags using tunnel diodes," in *The 25th Annual International Conference on Mobile Computing and Networking*, ser. MobiCom '19. New York, NY, USA: Association for Computing Machinery, 2019. [Online]. Available: https://doi.org/10.1145/3300061.3345451

[42] M. V. Thayyil, A. Figueroa, N. Joram, and F. Ellinger, "Integrated super-regenerative amplifier based 24 ghz fmcw radar active reflector tags for joint ranging and communication," *IET Radar, Sonar & Navigation*, vol. 17, no. 8, pp. 1196–1212, 2023.

[43] A. Okba, A. Takacs, H. Aubert, S. Charlot, and P.-F. Calmon, "Multiband rectenna for microwave applications," *Comptes Rendus Physique*, vol. 18, no. 2, pp. 107–117, 2017, energy and radiosciences. [Online]. Available: https://www.sciencedirect.com/science/article/pii/S1631070516301852

[44] S. Ladan, A. B. Guntupalli, and K. Wu, "A high-efficiency 24 ghz rectenna development towards millimeter-wave energy harvesting and wireless power transmission," *IEEE Transactions on Circuits and Systems I: Regular Papers*, vol. 61, no. 12, pp. 3358–3366, 2014.

[45] A. Eid, J. Hester, and M. M. Tentzeris, "Extending the range of 5g energy transfer: Towards the wireless power grid," in *2022 16th European Conference on Antennas and Propagation (EuCAP)*, 2022, pp. 1–4.

[46] A. Eid, J. Hester, and M. Tentzeris, "5g as a wireless power grid," *Scientific Reports*, 01 2021.

[47] E. Shaulov, T. Elazar, and E. Socher, "A high sensitivity cmos rectifier for 5g mm-wave energy harvesting," *IEEE Transactions on Circuits and Systems I: Regular Papers*, vol. 71, no. 7, pp. 3041–3049, 2024.

[48] T. Elazar, E. Shaulov, and E. Socher, "Analysis of mm-wave cmos rectifiers and ka-band implementation," *IEEE Transactions on Microwave Theory and Techniques*, vol. 71, no. 6, pp. 2758–2768, 2023.